\documentclass{ifacconf}
\usepackage{graphicx}      
\usepackage{natbib}        
\usepackage{amsmath,amssymb,amsfonts}
\usepackage{mathrsfs} 
\usepackage{algorithm}  
\usepackage{balance}

\usepackage{algorithmic}
	
	\usepackage{textcomp}
	\usepackage{bm}
	\usepackage{makecell}
	\usepackage{mathdots}
	\usepackage{pgfplots}
	
	\usepackage{enumitem}
	\setlist{nolistsep}
	\pgfplotsset{compat=1.16}

	\newcommand{\Rb}{{\mathbb{R}}}
	\newcommand{\Cb}{{\mathbb{C}}}
	
	\newcommand{\Zb}{{\mathbb{Z}}}
	
	\newcommand{\Ob}{{\mathbb{O}}}
	\newcommand{\Ab}{{\mathbb{A}}}

	\newcommand{\Gc}{{\mathcal{G}}}

	\newcommand{\Fc}{{\mathcal{F}}}

	\newcommand{\Qc}{{\mathcal{Q}}} 
	\newcommand{\Ac}{{\mathcal{A}}}
	\newcommand{\Bc}{{\mathcal{B}}}
	\newcommand{\Cc}{{\mathcal{C}}}
	\newcommand{\Ic}{{\mathcal{I}}}

	\newcommand{\Sc}{{\mathcal{S}}}

	\newcommand{\Jc}{{\mathcal{J}}}

	\newcommand{\ra}{{\rightarrow}}
	\newcommand{\ift}{{\infty}}

	\newcommand{\tAi}{\tilde{A}_i}
	\newcommand{\tCi}{\tilde{C}_i}

	\DeclareMathOperator{\rs}{{rowspan}}
	
	\DeclareMathOperator{\Span}{{span}}
	\DeclareMathOperator{\rank}{{rank}}
	
	\DeclareMathOperator{\diag}{{diag}}
	\DeclareMathOperator{\supp}{{supp}}

	\DeclareMathOperator{\med}{{med}}
	\DeclareMathOperator{\remain}{{rem}}
	
	\theoremstyle{plain}
	\newtheorem{remark}{\textbf{Remark}}
	\newtheorem{assumption}{\textbf{Assumption}}
	\newtheorem{definition}{\textbf{Definition}}

\begin{document}
		\begin{frontmatter}
			
			\title{Secure State Estimation against Sparse Attacks on a Time-varying Set of Sensors\thanksref{footnoteinfo}} 
			
			\thanks[footnoteinfo]{This work is supported by National Natural Science Foundation of China under grant No. 62273196 and 62192752, and the Swedish Research Council, the Swedish Foundation for Strategic Research, and the Knut and Alice Wallenberg Foundation, Sweden. It also received funding by the European Union's Horizon Research and Innovation Programme under Marie Sk\l{}odowska-Curie grant agreement No. 101062523.}
			
			\author[thu]{Zishuo Li} 
			\author[kth,mit]{Muhammad Umar B. Niazi} 
			\author[kth]{Changxin Liu}
			\author[thu]{Yilin Mo}
			\author[kth]{Karl H. Johansson}
			
			\address[thu]{Department of Automation, Tsinghua University, Beijing, 100084, China (e-mail: \rm \texttt{lizs19@mails.tsinghua.edu.cn}, \texttt{ylmo@tsinghua.edu.cn})}
			\address[kth]{Division of Decision and Control Systems, Digital Futures, School of Electrical Engineering and Computer Science, KTH Royal Institute of Technology, Stockholm SE-100 44, Sweden (e-mail: \rm \texttt{changxin@kth.se}, \texttt{kallej@kth.se})}
			\address[mit]{Laboratory for Information and Decision Systems, Massachusetts Institute of Technology, 77 Massachusetts Avenue, Cambridge, MA 02139, USA (e-mail: \rm \texttt{niazi@mit.edu})}
			
			\begin{abstract}                
				This paper studies the problem of secure state estimation of a linear time-invariant (LTI) system with bounded noise in the presence of sparse attacks on an unknown, time-varying set of sensors. In other words, at each time, the attacker has the freedom to choose an arbitrary set of no more that $p$ sensors and manipulate their measurements without restraint. To this end, we propose a secure state estimation scheme and guarantee a bounded estimation error subject to $2p$-sparse observability and a mild, technical assumption that the system matrix has no degenerate eigenvalues. The proposed scheme comprises a design of decentralized observer for each sensor based on the local observable subspace decomposition. At each time step, the local estimates of sensors are fused by a median operator to obtain a secure estimation, which is then followed by a local detection-and-resetting process of the decentralized observers. The estimation error is shown to be upper-bounded by a constant which is determined only by the system parameters and noise magnitudes. Moreover, we design the detector threshold to ensure that the benign sensors never trigger the detector. The efficacy of the proposed algorithm is demonstrated by its application on a benchmark example of IEEE 14-bus system.
			\end{abstract}
			
		\end{frontmatter}
		
		\section{Introduction}
		Cyber physical systems tend to be vulnerable to intelligently devised malicious attacks because of the interconnectedness of remote sensors and unprotected cyber-infrastructures operating physical plants. This has caused significant concern in the research community \citep{cardenas2009challenges}, who have highlighted the dire need for resilient/secure estimation and control algorithms. 
		
		In this paper, we consider the secure state estimation of an LTI system in the presence of attacks on an unknown subset of sensors, which may vary with time. However, the attacker is assumed to possess limited resources and, at each time instant, can corrupt at most $p$ out of total $m$ sensors by injecting arbitrary signals to their measurements.
		A milder scenario widely considered in the literature assumes that the attacked subset of sensors is fixed with respect to time. Thus, it is usually easier to detect attacks in such a scenario. The problem becomes significantly challenging when the attacked subset is time-varying, and, in this regard, the method proposed in this paper is quite fundamental.
		
		There are three main research directions for tackling the problem of secure state estimation with fixed subset of attacked sensors. The first is the sliding window method that formulates the state estimation task as a batch optimization problem by considering the past measurements of finite time window.
		Under certain observability conditions, \cite{FawziTAC2014} and \cite{ShoukryTAC2016} propose a method to minimize the norm of the residue signal for obtaining a secure state estimate. However, searching for the minimizer is of combinatorial nature because the corrupted sensor set is unknown. \cite{Shoukry2017} employ Satisfiability Modulo Theory (SMT) to reduce the size of attacked sensor subset by pruning the optimization process. Nonetheless, sliding window method discards history information by ignoring the measurements out of the considered window. 
		The second method is based on estimator switching that maintains multiple parallel estimators based on measurements from sensor subsets with cardinality $m-p$. By assuming that the corrupted subset of sensors is fixed, this method guarantees the existence of at least one subset of sensors unaffected by the attack. \cite{Yorie2018TAC} devise a detection scheme to select all trusted subsets of sensors and fuse the corresponding local estimates for secure estimation. However, maintaining a combinatorial number of estimators simultaneously introduces heavy computational and storage burden. \cite{Mishra2017TCNS} replace the exhausted search of sensor subsets by SMT based searching, which can improve the computational load to a great degree. \cite{luTAC2019} resort to set cover approach to reduce the number of subset candidates by at least half. \cite{Yang2018TAC} adopt a sequential switching mechanism where the algorithm switches from a malicious subset to a trusted one until it converges to the benign subset of subsets.
		The third method is based on local decomposition-fusion that solves a combinatorial complexity problem by designing $m$ decentralized estimators based on centralized estimator and fusing the local estimates by a convex optimization problem \citep{liuxinghuaTAC2020,zishuo_RNC2021}. The optimal trade-off between the security and estimation accuracy can be achieved by carefully choosing the optimization parameters.
		
		Allowing attacks on a time-varying subset of sensors is more practical than the setup where the corrupted subset is fixed.
		However, the problem becomes significantly challenging in the following ways: (1)~Historical knowledge on corrupted sensors becomes useless because the attacker can arbitrarily choose a totally different set at each time. (2)~Computational complexity increases since the corrupted subset of sensors needs be searched at every time step. 
		Because of the aforementioned challenges, the secure state estimation problem under time-varying attacked set is not well-explored. In this regard, \cite{he2022TAC} propose a saturation-gain filter where the gain of the innovation term is saturated if the residue signal exceeds the prescribed threshold. However, this comes at the cost of a restrictive assumption that the system is 1-step $2p$-sparse observable, i.e.\footnote{The matrix $C_{\Ic\setminus\Cc}$ represents the matrix composed of rows of $C$ with row index in $\Ic\setminus\Cc$.}, $\rank(C_{\Ic\setminus\Cc})=n, \forall |\Cc|=2p$.
		\cite{An_22_TC_timevarying} proposed an switching based non-linear observer that asymptotically converges to a neighborhood of the true system state in the presence of time-varying sensor attacks. In this paper, we propose a decentralized observer scheme that has bounded worst case estimation error in spite of time-varying attacked set, under $2p$-sparse observability. Our contribution can be summarized as follows:
		
		\begin{itemize}[left=0pt]
			\item \textbf{Estimation performance:} We derive explicit estimation error bounds under $p$-sparse attack on an arbitrary time-varying subset of sensors. 
			\item \textbf{Observability requirement:} Our method requires the usual $2p$-sparse observability condition. This assumption is fundamental in the literature that assumes fixed subset of attacked sensors, and we show that no additional requirement is needed in the time-varying case.
			\item \textbf{Low online computational complexity:} As we employ reduced-order decentralized estimator design, we need less online computational and storage resources compared to \cite{liuxinghuaTAC2020}.
			\item \textbf{Low offline computational complexity:} The $2p$-sparse observability requirement can be verified in polynomial time with respect to the number of sensors and the dimension of the system, under the assumption that the system matrix has no degenerate eigenvalues. 
		\end{itemize}
		
		The rest of the paper is organized as follows. We formulate the problem of secure state estimation in Section~\ref{sec:problem_formulation} and describe the local observable subspace decomposition in Section~\ref{sec:pre}. Then, in Section~\ref{sec:central_est}, we present the design of decentralized observers and their fusion scheme. In the same section, we present the overall algorithm and provide the performance guarantee (Theorem~\ref{th:final}). In Section \ref{sec:simulation}, a numerical example is demonstrated to corroborate the performance of the proposed method. Finally, Section \ref{sec:conclusion} concludes the paper with a future outlook.
		
		\textit{Notations:} The set of positive integers is denoted by $\Zb^+$.
		The $n$-dimensional real and complex vector spaces are denoted by $\Rb^n$ and $\Cb^n$, respectively.
		Dimension of a linear vector space is denoted as $\dim(\cdot)$. 
		The notation $[x]_j$ specifies the $j$-th entry of a vector $x$.
		By $\rs(A)$, we represent the linear span of the rows of matrix $A$.
		Conjugate transpose of a matrix $A$ is $A'$.
		We use $\|\cdot\|_p$ to denote the $p$-norm of a vector or the induced $p$-norm of a matrix, which should be clear from the context.
		Finally, $\sigma_{\max}(\cdot)$ denotes the maximum singular value of a matrix.

		\section{Problem Formulation}\label{sec:problem_formulation}
		Consider the discrete-time LTI system described by
		\begin{align}
			x(k+1)&=A x(k)+w(k)  \label{eq:system} \\
			y(k)&=C x(k)+v(k)+a(k) \label{eq:y_i_def}
		\end{align}
		where $x(k) \in \mathbb{R}^{n}$ is the system's state, $w(k)\in\mathbb{R}^{n},v(k)\in\Rb^{m}$ are the bounded process noise and measurement noise, respectively, with $\|w(k)\|_2 \leq \Bc_w$ and $\|v(k)\|_2 \leq \Bc_v$, for every $k\in\{0\}\cup\Zb^+$. 
		The vector $y(k)\in \mathbb{R}^{m}$ is the collection of measurements from all $m$ sensors, and $i$-th entry $y_i(k)$ is the measurement from sensor~$i$.
		The vector $a(k)$ denotes the sensor attack by an adversary and $a_i(k)$ represents the attack signal (malicious data) injected into sensor~$i$'s measurement at time $k$. We denote the set of sensors as $\Ic=\{1,2,\cdots,m\}$. 
		We consider the scenario where a time-varying subset of sensors is compromised by a malicious adversary. 
		Let the support of vector $a\in\Rb^{m}$ be 
		\[
		\supp(a)\triangleq \left\{i| 1\leq i\leq m , a_i\neq0 \right\}.
		\]
		
		\begin{definition}[Time-varying sparse attack]
			\label{def:attack}
			The attack $a(k)$ is said to be a \textit{time-varying $p$-sparse attack} if the compromised set $\Cc(k)\triangleq \supp\left\{a(k)\right\}$ satisfies $|\Cc(k)| \leq p$ for all $k$. The set of all \textit{admissible} time-varying $p$-sparse attacks is then defined as $\Ab_p$.  \quad $\diamond$
		\end{definition}
		
		It is important to emphasize that the estimation scheme has no direct information about $\Cc(k)$, but it knows that the maximum number of corrupted sensors is $p$.
		
		Another notion that needs to be defined is that of sparse observability \citep{ShoukryTAC2016}, which characterizes the redundancy of system's observability.

		\begin{definition}[Sparse observability]
			\label{def:sparse_obs}
			The pair $(A,C)$ is said to be \textit{$s$-sparse observable} if the pair $(A,C_{\Ic\setminus\Cc})$ is observable\footnote{The matrix $C_{\Ic\setminus\Cc}$ represents the matrix composed of rows of $C$ with row index in $\Ic\setminus\Cc$.} for any subset of sensors $\Cc\subset\Ic$ with cardinality $|\Cc| = s$.  \quad $\diamond$
		\end{definition}

		
		The secure state estimation problem aims at recovering the system's state $x(k)$ at every time $k$ with uniformly bounded error through sensor measurements, which might have been partly manipulated.
		
		\begin{definition}[Secure estimator]\label{def:resi}
			Denote $\hat{x}(k)$ as the estimate at time $k$. Then, an estimator is said to be \textit{secure} against time-varying $p$-sparse attack if there exists a constant $\Bc_e\geq 0$ determined only by the noise magnitudes $\Bc_w,\Bc_v$ and system matrices $A,C$ such that, for all time $k$ and all attacks $a\in\Ab_p$, 
			$$\sup_{\|w\|_2\leq \Bc_w,\|v\|_2\leq \Bc_v} \|\hat{x}(k)-x(k)\|_2 \leq \Bc_e . \quad \diamond$$
		\end{definition}
		
		In this paper, we propose a secure state estimation scheme against time-varying $p$-sparse attacks, which only requires $2p$-sparse observability. This condition is less restrictive than the $1$-step $2p$-sparse observability~\citep{he2022TAC}, and is necessary in the sense of exact state recovery when the system is noise free~\citep{ShoukryTAC2016}.  
		
		\section{Preliminaries on Local Observable Subspace Decomposition}\label{sec:pre}
		
		Before introducing our design of decentralized observers, we need some preliminaries on the decomposition of observable space into local observable subspace.
		In order to better represent and denote the structure of the system observable space, we consider the following assumption on the system matrix $A$.  
		\begin{assumption}\label{as:sepA}
			All the eigenvalues of matrix $A$ have geometric multiplicity 1. Without loss of generality, we can assume that $A$ is in the Jordan canonical form. \quad $\diamond$
		\end{assumption}
		\begin{remark}
			If some eigenvalues of $A$ have geometric multiplicity greater than 1, then the observability structure becomes complicated, which makes expressing the observable space intractable, and validating the sparse observability (Definition~\ref{def:sparse_obs}) is NP-hard \citep{mao2021computational}. Therefore, for secure state estimation, it is quite common to assume that $A$ does not have degenerate eigenvalues \citep{shim2020TAC_med}. \quad $\diamond$
		\end{remark}
		The following assumption is needed for observability redundancy to resiliently recover system state.
		\begin{assumption}\label{as:sparse_obs}
			The system $(A,C)$ is $2p$-sparse observable. \quad $\diamond$
		\end{assumption}
		
		Recall the index set of sensors $\Ic\triangleq \{1,2,\cdots,m\}$. Define the index set of state entries as $\Jc\triangleq \{1,2,\cdots,n\}$ and the observability matrix with respect to sensor~$i$ as
		\begin{equation}\label{eq:def_O}
			O_{i} \triangleq \begin{bmatrix}
			C_{i} ^\top&
			\left(C_{i} A\right){^\top}&\cdots&\left(C_{i} A^{n-1}\right){^\top}
		\end{bmatrix}^\top .
		\end{equation}
		Then, the observable subspace of sensor~$i$ is
		\begin{equation*}
			\Ob_i\triangleq\rs(O_i)=\Span\left(C'_i,\left(C_{i} A\right){'},\cdots,\left(C_{i} A^{n-1}\right){'}\right).
		\end{equation*}
		All the states that can be observed by the measurements from sensor~$i$ belong to the linear vector space $\Ob_i$. The dimension of $\Ob_i$ is denoted as $n_i$.
		The system's observable space is given by $\Ob\triangleq \bigcup_{i\in\Ic} \Ob_i$. 
		Define $\mathbf{e}_j$ as the $n$-dimensional canonical basis vector with 1 on the $j$-th entry and 0 on all the other entries, then the following theorem characterizes the transformation between $\Ob$ and $\Ob_i$.
		\begin{thm}\label{th:decomp_obs}
			For each sensor $i$, there exists a linear projection $\Ob\ra \Ob_i$ represented by matrix $H_i$, where $H_i$ is an $n_i\times n$ matrix, such that, for any arbitrary $x\in\Ob$, $H'_i H_i x=x$ if $x\in\Ob_i$, and $H'_i H_i x=\bm{0}$ if $x\in\Ob\setminus\Ob_i$.
			
			Moreover, if Assumption \ref{as:sepA} holds, $H_i$ can be constructed as the following $n_i\times n$ matrix:
			\begin{align}\label{eq:defH}
				H_i = 
				\begin{bmatrix}
					\mathbf{e}_{j_1}& \mathbf{e}_{j_2}& \cdots &\mathbf{e}_{j_{n_i}}
				\end{bmatrix}'
			\end{align}
			where $\{j_1,\cdots,j_{n_i}\}=\Qc_i$ with
			\begin{equation}\label{eq:def_Ec}
				\Qc_i\triangleq \{j\in\Jc\ |\ O_i \mathbf{e}_j\neq \mathbf{0} \}.
			\end{equation}
		\end{thm}
		\begin{pf}
			See Appendix A. \qed
		\end{pf}
		Theorem \ref{th:decomp_obs} provides an explicit formulation of the transformation from $\Ob$ to $\Ob_i$, and will facilitate our design of decentralized observers in the next section. 
		
		Define
		\begin{align}
			\tilde{A_i}\triangleq H_i A H'_i \in\Cb^{n_i\times n_i},\quad \tilde{C_i}\triangleq C_i H'_i \in\Cb^{1\times n_i}
		\end{align} 
		where $C_i$ is $i$-th row of matrix $C$.
		The following properties of the transformation matrix $H_i$ are relevant.
		\begin{lem}\label{lm:subspace}
			If Assumption \ref{as:sepA} is satisfied, then the following holds for all $i\in\Ic$:
			\begin{align}
				\tAi H_i&=H_i A \label{eq:HAH}\\
				\tCi H_i&=C_i.\label{eq:CHH}
			\end{align}
			Moreover, the pair $(\tAi,\tCi)$ is observable.
		\end{lem}
		\begin{pf}
			See Appendix B. \qed
		\end{pf}

		
		For the implications of such a transformation $H_i$, consider a simple illustrative example where
		$$A=\begin{bmatrix}
			\lambda_1 &0&0\\
			0& \lambda_2 & 1\\
			0&0&\lambda_2
		\end{bmatrix}, \quad 
		C=\begin{bmatrix}
			1 &0&0\\
			0&1&0
		\end{bmatrix}.$$
		Then, notice that sensor 1 can only observe state 1, while sensor 2 can observe both state 2 and 3. Thus, we have $\Qc_1=\{1\}$ and $\Qc_2=\{2,3\}$, and 
		$$H_1=\begin{bmatrix}
			1&0&0
		\end{bmatrix}, \quad H_2=\begin{bmatrix}
			0&1&0\\ 0&0&1
		\end{bmatrix}$$
		with
		$$\tilde{A}_1=\lambda_1,\; \tilde{C}_1=1,\quad \tilde{A}_2=\begin{bmatrix}
			\lambda_2 & 1\\
			0&\lambda_2
		\end{bmatrix}, \; \tilde{C}_2=\begin{bmatrix}
			1 &0
		\end{bmatrix}.$$
		It is straightforward to verify that equations \eqref{eq:HAH} and \eqref{eq:CHH} hold.
		By this example, it is clear that $\tAi,\tCi$ are truncated matrices corresponding to the observable subspace of sensor $i$.

		\section{Secure Estimation Design}\label{sec:central_est}
		
		In this section, we propose a design of secure estimation based on decentralized local state observers and local detectors as well as a resilient centralized fusing scheme by median operators. 
		
		
		\subsection{Secure Estimation Algorithm}
		
\begin{figure}[!ht]
			\centering
			\usetikzlibrary{arrows}
\begin{tikzpicture}[auto, node distance=1.8cm,>=latex', font=\small]
	
	\draw (-0.1,1.7) rectangle (0.7,1.1);
	\node at (0.3,1.5) {\scriptsize step} ; 
	\node at (0.3,1.3) {\scriptsize (iii)} ; 
	
	\draw [rounded corners](-4.5,2.3) rectangle (-3.1,1.7);
	\node at (-3.8,2) {$y_1(k-1)$};
	\draw (-3.1,2) -- (-2.3,2);
	\draw [->](-2.3,2) -- (-2.3,1.6);

	\draw (-2.8,1.6) rectangle (-1.8,1.2);
	\node at (-2.3,1.4) {step(i)};

	\node at (-5.1,1.4) {$\cdots$};
	\draw [->] (-4.8,1.4) -- (-4.5,1.4);
	\node at (-3.8,1.4) {$\eta^+_1(k-1)$};
	\draw [->] (-3.1,1.4) -- (-2.8,1.4);
	\node at (-1.0,1.4){$\eta_1(k)$};
	\draw [->] (-0.6,1.4) -- (-0.1,1.4);
	\draw [->] (0.7,1.4) -- (1.0,1.4);
	\node at (1.4,1.4){$\eta^+_1(k)$};
	\draw [->] (1.8,1.4) -- (2.1,1.4);
	
	\draw [->] (-1.8,1.4) -- (-1.4,1.4);

		\node at (2.4,-2) {$\cdots$};
		\node at (2.4,0) {$\cdots$};
		\node at (2.4,1.4) {$\cdots$};

	\draw (-0.1,-0.3) rectangle (0.7,0.3);
	\node at (0.3,0.1) {\scriptsize step} ; 
	\node at (0.3,-0.1) {\scriptsize (iii)} ; 
	
	\draw [rounded corners](-4.5,0.9) rectangle (-3.1,0.3);
	\node at (-3.8,0.6) {$y_2(k-1)$};
	\draw (-3.1,0.6) -- (-2.3,0.6);
	\draw [->] (-2.3,0.6) -- (-2.3,0.2);
	
	\draw (-2.8,0.2) rectangle (-1.8,-0.2);
	\node at (-2.3,0) {step(i)};
	\draw [->] (-1.8,0) -- (-1.4,0);

	\node at (-5.1,0) {$\cdots$};
	\draw [->] (-4.8,0) -- (-4.5,0);
	\node at (-3.8,0) {$\eta^+_2(k-1)$};
	\draw [->] (-3.1,0) -- (-2.8,0);
	\node at (-1.0,0){$\eta_2(k)$};
	\draw [->] (-0.6,0) -- (-0.0,0);
	\draw [->] (0.6,0) -- (1.0,0);
	\node at (1.4,0){$\eta^+_2(k)$};
	\draw [->] (1.8,0) -- (2.1,0);

	\draw (-0.1,-2.3) rectangle (0.7,-1.7);
	\node at (0.3,-1.9) {\scriptsize step} ; 
	\node at (0.3,-2.1) {\scriptsize (iii)} ;

	\draw [rounded corners](-4.5,-1.7) rectangle (-3.1,-1.1);
	\node at (-3.8,-1.4) {$y_m(k-1)$};
	\draw (-3.1,-1.4) -- (-2.3,-1.4);
	\draw [->] (-2.3,-1.4) -- (-2.3,-1.8);

	\draw (-2.8,-1.8) rectangle (-1.8,-2.2);
	\node at (-2.3,-2) {step(i)};
	\draw [->] (-1.8,-2) -- (-1.5,-2);

	\node at (-5.1,-2) {$\cdots$};
	\draw [->] (-4.8,-2) -- (-4.5,-2);
	\node at (-3.8,-2) {$\eta^+_m(k-1)$};
	\draw [->] (-3.1,-2) -- (-2.8,-2);
	\node at (-1.0,-2.0){$\eta_m(k)$};
	\draw [->] (-0.6,-2.0) -- (-0.0,-2.0);
	\draw [->] (0.6,-2.0) -- (1.0,-2.0);
	\node at (1.4,-2.0){$\eta^+_m(k)$};
	\draw [->] (1.8,-2.0) -- (2.1,-2);

	
	\node at (-3.8,-0.6) {$\vdots$};
	\node at (-1,-0.6){$\vdots$};
	\node at (0.3,-0.6){$\vdots$};
	\node at (1.3,-0.6){$\vdots$};

	\draw [rounded corners, blue, dashed, line width=1.0pt](-0.5,1.8) rectangle (-1.6,-2.3);
	\draw [->][blue] (-1.,-2.3) -- (-1,-2.7);
	\node at (-2.1,-2.8) {\color{blue} \tiny Secure };
	\node at (-2.1,-3) {\color{blue} \tiny Fusion };
	
	\draw [blue, line width=0.5pt](-0.4,-2.7) rectangle (-1.6,-3.1);
	
	\node at (-1,-2.9) {step(ii)};
	
	\draw [->][blue] (-1,-3.1) -- (-1,-3.4);
	
	\node at (-1,-3.6) {$\hat{x}(k)$};
	\draw [orange] (-0.6,-3.6) -- (0.3,-3.6);
	\draw [->][orange] (0.3,-3.6) -- (0.3,-2.5);
	
	\node at (0.9,-2.8) {\color{orange} \tiny Detection };
	\node at (0.9,-3) {\color{orange} \tiny and Reset };

	\draw [rounded corners, orange, dashed, line width=1.0pt](0.9,1.9) rectangle (-0.3,-2.5);

\end{tikzpicture}
			\caption{The information flow of our proposed estimation scheme. 
} \label{fig:flow}
		\end{figure}
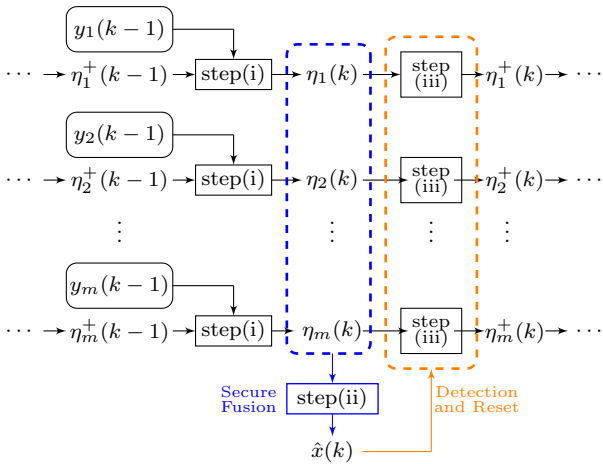

		Our proposed algorithm is summarized in the following Algorithm \ref{al:cen} and Fig. \ref{fig:flow}. At each time instant, the algorithm is composed of three steps, i.e., (i) local observer state update; (ii) secure fusion; (iii) local detection and reset. As shown in Fig.\ref{fig:flow},  step (i) and (iii) are performed in a decentralized manner, and step (ii) fuses the local information to obtain a secure estimation by the median operator.
		
		In step (i) (line 4 in the algorithm), the local state $\eta_i^+(k-1)$ is updated to $\eta_i(k)$ using sensor measurement $y_i(k-1)$ by \eqref{eq:update_local}. In step (ii)  (line 5 in the algorithm), secure estimation $\hat{x}(k)$ is obtained by fusing all the local states $\eta_i(k)$ using the median operation in \eqref{eq:x^*=med}, where $\Fc_j$ is designed as the index set of sensors that can observe state $j$ as follows:
		\begin{align}
			\Fc_j\triangleq \{i\in \Ic \ |\ O_i \mathbf{e}_j\neq \mathbf{0} \}.
		\end{align}
		And the median operator is defined as 
		\begin{align}\label{eq:def_med}
			\med(z_i,i\in\Sc)\triangleq f_{(|\Sc|+1)/2} (z_i,i\in\Sc),
		\end{align}
		where operator $f_{s}(z_i,i\in\Sc)$ equals to the $s$-th smallest element in the set $\{z_i|i\in\Sc\}$.
		In step (iii) (lines 6-10 in the algorithm), by local detection and resetting, state $\eta_i(k)$ is reset as $\eta_i^+(k)$ for the usage of next time step.

		\begin{algorithm}[ht]
			\caption{Secure Estimation at sensor $i$ }\label{al:cen}
			\begin{algorithmic}[1]
				\STATE {Offline parameters: matrices $H_i$, constant $\gamma$}
				\STATE Initialize  $\eta^+_i(0):=H_i\hat{x}(0)$
				\FOR {every time index $k\in\Zb^+$}
				\STATE Update $\eta_i(k)$ by 
				\begin{align}\label{eq:update_local}
					\eta_i(k):=\left(\tAi-L_i \tCi\right)\eta^+_i(k-1)+L_i y_i(k-1)
				\end{align}
				
				\STATE Calculate each entry of $\hat{x}(k)$ by
				\begin{align}\label{eq:x^*=med}
					\left[\hat{x}(k)\right]_j={\med} \left\{\left[H_i^\top \eta_i[k]\right]_j, i\in\Fc_j \right\},\ \forall j \in\Jc
				\end{align}	 
				
				\IF {$\|\eta_i(k)-H_i\hat{x}(k) \|_2 >(\sqrt{n_i}+1)\gamma$}
				\STATE {$\eta^+_i(k):=H_i \hat{x}(k)$}
				\ELSE
				\STATE {$\eta^+_i(k):=\eta_i(k)$}
				\ENDIF
				\ENDFOR
			\end{algorithmic}
		\end{algorithm}
		
		Even though we consider centralized estimation problems, the algorithm is designed in a decentralized manner to isolated the influence of corrupted sensors, i.e., compromised $y_i(k)$ will never influence the value of $\eta_{j}(k),j\neq i$. In this way, the combinatorial complex problem of finding the corrupted sensor set is simplified as a resilient fusion problem which can be solved in a computationally easy way, i.e., taking median. 
		The security of our proposed Algorithm \ref{al:cen} is proved in the next subsection.
		
		
		Moreover, since the dimension of decentralized observer is reduced to $n_i\leq n$, where $n_i$ is the dimension of the subsystem $(\tAi,\tCi)$, maintaining such an observer requires less computational and storage resources compared to $n$-dimensional observer in the previous works \citep{liuxinghuaTAC2020,zishuo_RNC2021}, especially in large scale systems where $n_i<<n$.

		\subsection{Security of the Proposed Algorithm}
		In this subsection, we prove the security of the algorithm by mathematical induction.
		
		Define the set of sensors that has estimation error smaller than $\gamma$ at time $k$ as ``accurate" sensor set $\Ac(k)$:
		\begin{align*}
			\Ac(k)\triangleq \{\ i\in\Ic\ \left| \ \|\eta_i(k)-H_ix(k)\|_{2}\leq \gamma \right. \} .
		\end{align*}
		Define good sensor set as those free of attack at time $k$, i.e., $\Gc(k)\triangleq\Ic\setminus \Cc(k)$. We have the following theorem providing the effectiveness of local detection/resetting and resilient fusion algorithm.
		\begin{thm}\label{th:induction_pre}
			Suppose Assumption \ref{as:sepA} and \ref{as:sparse_obs} are satisfied. Moreover, assume for each sensor $i$, estimator gain $L_i$ and detector parameter $\gamma$ satisfy the following inequality:
			\begin{align}\label{eq:lambda_A-KC}
				\sigma_{\max}\left(\tAi-L_i\tCi\right)\leq \frac{\gamma-\Bc_w- \|L_i\|_2\Bc_v }{(2\sqrt{n_i}+1)\gamma}
			\end{align} with $\gamma\geq\Bc_w+ \|L_i\|_2\Bc_v$. 
			Then we have the following implication results:
			\begin{align*}
				|\Ac(k)|\geq m-p \Rightarrow \text{(s1)} \Rightarrow \text{(s2)} \Rightarrow \text{(s3)}
			\end{align*}
			where the three statements are:
			\begin{itemize}
				\item[(s1)] Resilient fusion by median operator at time $k$ is bounded:
				\begin{align}
					\|\hat{x}(k)-x(k)\|_\ift \leq \gamma.\label{eq:tildex_bounded }
				\end{align}
				\item[(s2)] Accurate sensor set covers good sensor set (good sensors always provide accurate local estimate):
				\begin{align*}
					\Ac(k+1)\supset \Gc(k).
				\end{align*}
				\item[(s3)] There will be at least same number of accurate local estimates at time $k+1$:
				\begin{align}
					|\Ac(k+1)|\geq m-p .\label{eq:Ac(k+1) }
				\end{align}
				
			\end{itemize}
			%
		\end{thm}
		\begin{pf}
		We first prove statement (s1). Define 
		\begin{align*}
			\xi_i(k)\triangleq H_i^\top \left( \eta_i(k)-H_ix(k) \right)
		\end{align*}
		Due to $|\Ac(k)|\geq m-p$, we have 
		\begin{align}\label{eq:xi_bounded}
		|\{i\in\Ic|\|\xi_i(k)\|_{\ift}<\gamma\}|\geq |\{i\in\Ic|\|\xi_i(k)\|_{2}<\gamma\}|\geq m-p.
		\end{align}
		Moreover, we have
		\begin{align*}
			[\hat{x}(k)]_j&=\med\left( [\xi_i(k)]_j+[H_i^\top H_i x(k)]_j, i\in\Fc_j \right)\\
			&=\med\left( [\xi_i(k)]_j+[x(k)]_j, i\in\Fc_j \right).
		\end{align*}
		Since the system is $2p$-sparse observable, we have $|\Fc_j|\geq2p+1,\forall j\in\Jc$. Due to the property of the median operator and inequality \eqref{eq:xi_bounded}, one obtains
		\begin{align*}
			\left| [\hat{x}(k)]_j-[x(k)]_j \right|\leq \gamma,
		\end{align*}
		and statement (s1) is proved.
		
		We prove statement (s2) by statement (s1) in the following, and statement (s3) automatically holds by statement (s2) since $|\Gc(k)|\geq m-p$ for $p$-sparse attacks.  
		After finishing the ``for" loop in Algorithm \ref{al:cen} at time $k$, there are two possible scenarios for the local estimate $\eta_i(k)$:
		\begin{itemize}
			\item[a)] $\eta_i(k)$ is reset to $\eta^+_i(k):=H_i \hat{x}(k)$,
			\item[b)] $\eta_i(k)$ does not trigger the detector and thus unaltered $\eta^+_i(k):=\eta_i(k)$, and satisfies $\|\eta^+_i(k)-H_i\hat{x}(k) \|_2 \leq (\sqrt{n_i}+1)\gamma$.
		\end{itemize}
		By Lemma~\ref{lm:subspace}, the following identity holds
		\begin{equation}\label{eq:sylvester}
			\left(\tAi-L_i \tCi\right)H_i=H_iA-L_iC_i.
		\end{equation}
		Thus, for good sensor $i\in\Gc(k)$,
		\begin{align*}
			& \eta_i(k+1)-H_ix(k+1)\\
			&=\left(\tAi-L_i \tCi\right)\eta^+_i(k)+L_i y_i(k) \\
			& \qquad\qquad\qquad\qquad\qquad -H_i\left[Ax(k)+w(k)\right]\\
			&=\left(\tAi-L_i \tCi\right)\eta^+_i(k)+L_i [C_ix(k)+v_i(k)]\\
			& \qquad\qquad\qquad\qquad\qquad -H_i\left[Ax(k)+w(k)\right]\\
			&=\left(\tAi-L_i \tCi\right) \left(\eta^+_i(k)-H_ix(k)\right)-H_{i} w(k) +L_i v_i(k) 
		\end{align*}
		where the last equality comes from \eqref{eq:sylvester}. 
		
		We prove statement (s2) for both scenario a) and b). 
		For scenario a) and good sensor $i\in\Gc$, we have
		\begin{align*}
			&\|\eta_i(k+1)-H_i {x}(k+1) \|_2\\
			&\quad =\|(\tAi-L_i\tCi) H_i(\hat{x}(k)-x(k)) -H_{i} w(k) +L_i v_i(k)\|_2 \\
			&\quad \leq \|\tAi-L_i\tCi\|_2\|H_i\|_2\|\hat{x}(k)-x(k)\|_2\\
			&\hspace*{12em} +\|H_i\|_2\Bc_w+\|L_i\|_2\Bc_v\\
			&\quad \leq \sigma_{\max}(\tAi-L_i\tCi) \sqrt{n_i}\gamma+\Bc_w+\|L_i\|_2\Bc_v ,
		\end{align*}
		where we used the facts $\|\hat{x}(k)-x(k)\|_2\leq \sqrt{n_i}\|\hat{x}(k)-x(k)\|_\ift\leq\sqrt{n_i}\gamma$ and $\|H_i\|_2=1$ based on statement (s1) and definition of $H_i$ in equation \eqref{eq:defH}, respectively.
		As a result, inequality \eqref{eq:lambda_A-KC} implies that 
		\begin{align}
			&\|\eta_i(k+1)-H_i {x}(k+1) \|_{2}\notag \\
			&\quad \leq\sigma_{\max}(\tAi-L_i\tCi) \sqrt{n_i}\gamma+\Bc_w+\|L_i\|_2\Bc_v \notag \\
			&\quad \leq\frac{\sqrt{n_i}\gamma}{(2\sqrt{n_i}+1)\gamma}\left(\gamma-\Bc_w-\|L_i\|_2\Bc_v\right)+\Bc_w+\|L_i\|_2\Bc_v\notag \\
			&\quad \leq\gamma-\Bc_w-\|L_i\|_2\Bc_v +\Bc_w+\|L_i\|_2\Bc_v, \label{eq:leq_gamma_1}
		\end{align}
		which results in $\|\eta_i(k+1)-H_i {x}(k+1) \|_{2}\leq \gamma,i\in\Gc$ for scenario a).
		
		For scenario b), similarly
		\begin{align*}
			&\|\eta_i(k+1)-H_i {x}(k+1) \|_2\\
			&\quad =\|(\tAi-L_i\tCi) \left[ \eta_i(k)-H_i\hat{x}(k)+ H_i(\hat{x}(k)-x(k))\right] \\
			&\hspace*{12em} -H_{i} w(k) +L_i v_i(k)\|_2 \\
			&\quad \leq \|(\tAi-L_i\tCi)\|_2\left((\sqrt{n_i}+1)\gamma+\sqrt{n_i}\gamma\right)\\
			&\hspace*{12em} +\|H_i\|_2\Bc_w+\|L_i\|_2\Bc_v\\
			&\quad \leq \sigma_{\max}(\tAi-L_i\tCi) (2\sqrt{n_i}+1)\gamma+\Bc_w+\|L_i\|_2\Bc_v,
		\end{align*}
		where we used $\|\eta_i(k)-H_i\hat{x}(k)\|_2\leq (\sqrt{n_i}+1)\gamma$ because the detector is not triggered in scenario b).
		Then, following the same steps as in \eqref{eq:leq_gamma_1}, we see that \eqref{eq:lambda_A-KC} implies $\|\eta_i(k+1)-H_i {x}(k+1) \|_{2}\leq \gamma,i\in\Gc$ for scenario b).
		Combining the above results, statement (s2) is proved and the proof is finished. $\qed$
	\end{pf}
	
		\begin{remark}
			The statement (s3) means that the our proposed detection algorithm has no false alarm or type-I error. Thus, no good sensor will be ``wronged" and all benign local states $\eta_i(k+1),i\in\Gc(k)$, will be directly used in the fusion step at time $k+1$. This means that the detector threshold is suitable in a sense that no useful information is discarded unnecessarily. This property is crucial for maintaining observability redundancy. 
			On the other hand, the detection algorithm may have miss detection or type-II error. Nevertheless, our proposed median based fusion in \eqref{eq:x^*=med} can tolerate the error introduced by undetected corrupted local states and generate secure estimation as shown in statement (s1). \quad $\diamond$
		\end{remark}
		
		The intuition behind condition \eqref{eq:lambda_A-KC} is that, in order to achieve stable state estimate against time-varying attacked set, each local estimator~$i$ should be ``stable" enough so that it immediately stops triggering the detector after the attacker has left sensor~$i$. 
		
		We present a necessary and sufficient condition under which such $\gamma,L_i$ satisfying \eqref{eq:lambda_A-KC} always exist.
		\begin{prop}\label{prop:when_exist}
			One can always find $\gamma$ and $L_i$ such that \eqref{eq:lambda_A-KC} holds if and only if the following holds:
			$$\min_{L_i} \sigma_{\max}(\tAi-L_i\tCi)<\frac{1}{2\sqrt{n_i}+1},\quad \forall i\in\Ic.$$
		\end{prop}
		By reformulating \eqref{eq:lambda_A-KC}, one obtains that
		\begin{align}\label{eq:gamma_bound}
			\gamma\geq \frac{\Bc_w+\|L_i\|_2\Bc_v}{1-(2\sqrt{n_i}+1)\|\tAi-L_i\tCi\|_2}.
		\end{align}
		The proof of Proposition \ref{prop:when_exist} is straightforward from \eqref{eq:gamma_bound} and \eqref{eq:lambda_A-KC}, and is omitted due to space limit.

			As one may notice in Theorem \ref{th:induction_pre}, the result is provided in an induction-based way, i.e., $$|\Ac(k)|\geq m-p \Rightarrow |\Ac(k+1)|\geq m-p.$$ In the following, by Theorem \ref{th:induction_pre}, we prove that the security of our propose estimation is guaranteed under time-varying $p$-sparse attacks as long as some initial condition is satisfied. This is the main result of the paper.
			
			\begin{thm}\label{th:final}
				Suppose Assumption \ref{as:sepA}-\ref{as:sparse_obs} and the following conditions are satisfied:
				\begin{itemize}
					\item[(i)] Inequality \eqref{eq:lambda_A-KC} holds.
					\item[(ii)] Either the initial estimate satisfies $\|\hat{x}(0)-x(0)\|_\ift\leq \gamma$ or $|\Ac(0)|\geq m-p$.
				\end{itemize}
				Then, Algorithm \ref{al:cen} provides a secure state estimate $\hat{x}$ in the sense that
				\begin{align}\label{eq:final_bound}
					\|\hat{x}(k)-x(k)\|_\ift \leq \gamma,\quad \forall k\in\Zb^+
				\end{align} 
				against any arbitrary time-varying $p$-sparse attack. 
				Moreover, good sensors at time $k$ will not trigger the detector at time $k+1$, i.e., $\forall i\in\Gc(k)$,
				\begin{align}\label{eq:no_wronged}
					\|\eta_i(k+1)-H_i\hat{x}(k+1)\|_2\leq (\sqrt{n_i}+1)\gamma.
				\end{align}
			\end{thm}
			
			\begin{pf}
				We prove \eqref{eq:final_bound} by mathematical induction. First, the condition (iii) makes sure $\|\hat{x}(0)-x(0)\|_\ift\leq\gamma$. By the induction structure of the result in Theorem \ref{th:induction_pre}, we know that estimation error bound \eqref{eq:final_bound} holds for all $k\in\Zb^+$ from statement (s1) and inequality \eqref{eq:no_wronged} holds from statement (s2).
				\qed
			\end{pf}
			
			
			The result in \eqref{eq:final_bound} can be interpreted as that the estimation error $\|\hat{x}(k)-x(k)\|_\ift$ is not worse than initial estimation error $\|\hat{x}(0)-x(0)\|_\ift$ under a time-varying set of compromised sensors.
			From a theoretical point of view, the result is non-trivial when the system is unstable, since the state estimate by pure prediction $\hat{x}(k)=A^k \hat{x}(0)$ has diverging error. Thus, a secure estimation algorithm must incorporate useful information from partly manipulated measurements and correct the state estimate to maintain a bounded error. From the practical perspective, the initial time $k=0$ is seen as the time when the attack is launched and an estimator has stable estimation already. Our proposed algorithm would not worsen the estimation error upper bound compared to when there is no attack, as long as the bound does not exceed $\gamma$ .

			\section{Numerical Simulation}\label{sec:simulation}
			We apply our proposed estimation scheme on the IEEE 14-bus system, which is a benchmark example extensively used in the literature \citep{George2010TPS,liu2020TSG} for illustrating the performance of secure state estimation algorithms. 
			We adopt the continuous-time system dynamics as in the following equations \citep{AllenPowerbook2013}:
			\begin{align*}
				\dot{\theta}_{i}(t) &=\omega_{i}(t) \\
				\dot{\omega}_{i}(t) &=-\frac{1}{m_{i}}\left[D_{i} \omega_{i}(t)+\sum_{j \in \mathcal{N}_{i}} P_{t i e}^{i j}(t)- P_{i}(t)+w_{i}(t)\right]
			\end{align*}
			where $\theta_{i}(t)$ and $\omega_{i}(t)$ are the phase angle and angular frequency on bus $i$, respectively, 
			$m_i$ is the angular momentum of $i$, and $w_i$ is the process disturbance.
			The parameter $D_i$ is the load change sensitivity w.r.t. the frequency and is defined in \cite{AllenPowerbook2013} Section 10.3.
			The power flow between neighboring buses $i$ and $j$ is given by $P_{\rm tie}^{i j}(t)=-P_{t i e}^{j i}(t)=t_{i j}\left(\theta_{i}(t)-\theta_{j}(t)\right)$, where $t_{ij}$ is the inverse of resistance between bus $i$ and $j$. 
			The power $P_{i}(t)$ denotes the difference of mechanical power and power demand, which is known to the system operator. 
			The system is sampled at discrete times with sampling interval $T_s=0.01$ s. Every bus is equipped with four sensors (one electric power sensor, one phase sensor and two angular velocity sensors).
			The process noise $w_i$ and measurement noise $v_{i}$ are i.i.d. uniformly distributed random vectors normalized to satisfy noise bounds $\Bc_w=10^{-3},\Bc_v=10^{-2}$. 
			The attack is switching between all 14 buses' electrical power sensor at every time index $k$. The attacked bus index is switching as the following rule:
			\begin{align*}
				\makecell{\text{corrupted}\\  \text{bus index}}
				=\begin{cases}
					\{1,2,3,4\}, &\remain(k,3)=1\\
					\{5,6,7,8\}, &\remain(k,3)=2\\
					\{9,10,11,12,13,14\}, &\remain(k,3)=0
				\end{cases}
			\end{align*}
			where $\remain(k,3)$ is the remaining of $k$ divided by 3. For random signal attack, $a_i(k)$ is a random value uniformly distributed in interval $[-10,10]$. For slope signal attack, $a_i(k)=k/5$. The detector parameter $\gamma$ is chosen to be $0.5$.
			
			\begin{figure}[!htb]
				\centering
				\input{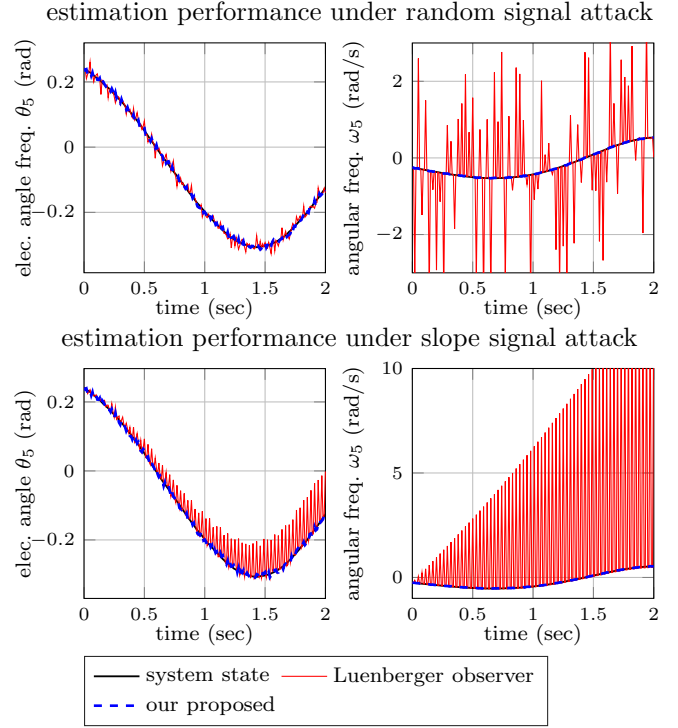}
				\caption{Estimation of states under slope signal attack and random signal attack on IEEE 14 system. The top two sub-figures illustrate the phase angle and angular velocity of bus 5 under random signal attack. The bottom two sub-figures illustrate those under slope signal attack.} \label{fig:IEEE14_est}
			\end{figure}
			Fig. \ref{fig:IEEE14_est} demonstrates the estimation performance on bus 5 under two kinds of attacks mentioned before. 
			The Luenberger observer used for the comparison is $
				\hat{x}(k+1)=A\hat{x}(k)+\sum_{i=1}^m H'_i L_i\left(y_i(k)-C_i\hat{x}(k)\right).
			$
			Since the electrical power measurement is directly related to angular frequency $\omega_i$, the estimation of $\omega_{5}$ is affected more significantly for insecure estimator such as Luenberger observer (shown in red line). In comparison, our proposed algorithm recovers the system state with small error (shown in blue dashed line).
			
			To better illustrate the effectiveness of our proposed detection-and-reset mechanism in Algorithm \ref{al:cen} line 6-10, we show the residue norm $\|\eta_i(k)-H_i \hat{x}(k)\|$ before and after reset in Fig. \ref{fig:IEEE14_detect}. Only first 30 samples are shown for presentation clearness. When the residue norm is larger than threshold $(\sqrt{n_i}+1)\gamma=2.081$ (blue triangle above red horizontal line), $\eta_i(k)$ is reset as $H_i\hat{x}(k)$ and the residue norm is decreased significantly (to be the blue circle). In Fig. \ref{fig:IEEE14_detect}, by this detection-and-reset mechanism, the residue holds stable and the result of Theorem \ref{th:final} is validated.
			\begin{figure}[!htb]
				\centering
				\input{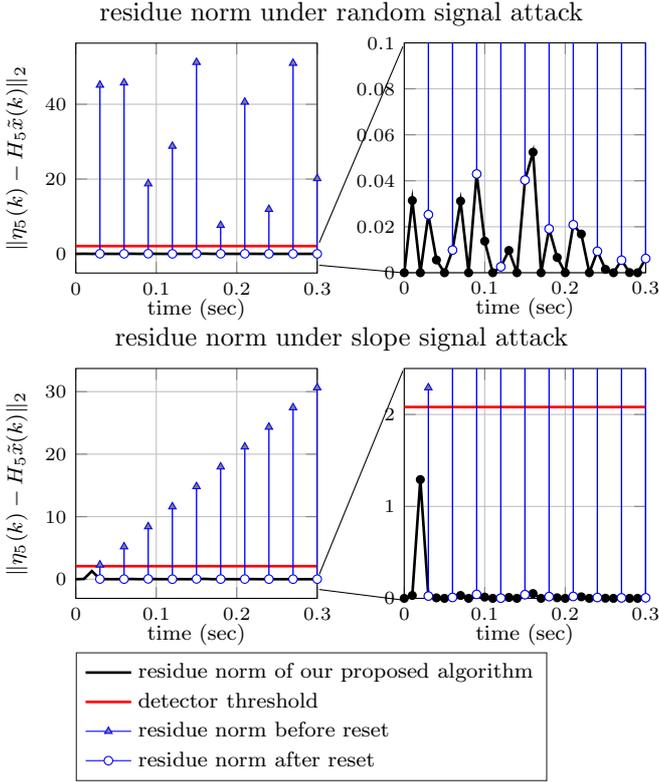}
				\caption{The residue norm $\|\eta_i(k)-H_i\hat{x}(k)\|_2$ before and after detection and resetting at bus $i=5$. Sub-figure on the right is the zoomed-in plot of the left. 
				} \label{fig:IEEE14_detect}
			\end{figure}
			
			\section{Concluding Remarks}\label{sec:conclusion}
			In summary, by designing local estimator based on observable space decomposition and a detection-resetting mechanism, we achieve secure estimation against sparse attack on time-varying sets. The update of local states $\eta_i$ as well as detection-resetting operations only relies on local information. Thus, it is straightforward to extend our method to a distributed sensor network by calculating \eqref{eq:x^*=med} in a distributed manner, which is left for future work. 
			Moreover, our design only applies for systems with dynamic matrix $A$ has no degenerate eigenvalues. One direction to close this gap is to study observable subspace decomposition on system where $A$ has degenerate eigenvalues, and design $H_i$ as the canonical basis vector group of $O_i$ which is not necessarily composed of $\mathbf{e}_j$. 
			Moreover, the fundamental limit of secure estimation under time-varying set attack is not-well studied to our best knowledge, and is left for future research.
			
			\bibliography{ifacconf.bib}            
			
			\appendix
			\section{Proof of Theorem \ref{th:decomp_obs}}   \label{ap:proof_th_decomp_obs}
			\begin{pf}
					We prove by constructing $H_i$.
					Since $\Ob_i$ is a $n_i$-dimensional linear subspace of $\Ob$, There exists a group of standard orthogonal basis vectors $\{b_1,b_2,\cdots,b_{d_i}\}$ such that $\Span(b_1,b_2,\cdots,b_{d_i})=\Ob_i$. Suppose $b_i$ are column vectors. Define matrix $$H_i\triangleq \begin{bmatrix}b_{1}&b_2& \cdots & b_{n_i}\end{bmatrix}\in\Cb^{n_i\times n}.$$

					For each $b_k,k\in\{1,2,\cdots,n_i\}$, 
					\begin{align*}
							H'_iH_ib_k= \sum_{j=1}^{n_i} b_jb'_j \cdot b_k=b_kb_k' b_k=b_k.
						\end{align*}
					The last equality comes from the fact that $b_j' b_k=\mathbf{0}$ if $j\neq k$ and $\|b_k\|_2=1$.
					
					Since each $x\in\Ob_i$ can be written as 
					$$x=\alpha_1 b_1 + \alpha_1 b_1 + \cdots + \alpha_{d_i} b_{d_i},$$
					we have $$H_i' H_ix=H_i' H_i\sum_{j=1}^{d_i} \alpha_j b_j=\sum_{j=1}^{d_i} \alpha_j b_j=x.$$
					
					We proceed to prove that when all the eigenvalues of $A$ has geometric multiplicity 1, $H_i$ is composed of canonical basis vectors. It is sufficient to prove that 
					\begin{equation}
							\Ob_i = \Span\left\{\mathbf{e}_j|j\in\Qc_i\right\}.
						\end{equation} 
					According to the definition of $\Qc_i$, $\Span\left\{\mathbf{e}_j|j\in\Qc_i\right\}=\Span\left\{\mathbf{e}_j|O_i{'} \mathbf{e}_j\neq \mathbf{0},j\in\Jc \right\}.$
					On the other hand, recalling to the definition of $\Ob_i$, we have
					\begin{align*}
							\Ob_i&\triangleq\Span\left\{C'_i,\left(C_{i} A\right){'},\cdots,\left(C_{i} A^{n-1}\right){'}\right\}\\
							&=\Span(O'_i\mathbf{e}_j,j\in\Jc).
						\end{align*}
					It is sufficient to prove that the following statements are equivalent.
					\begin{enumerate}
							\item[(i)] Vectors in set $\{O_i \mathbf{e}_j | O_i \mathbf{e}_j\neq \mathbf{0},j\in\Jc\}$ are linear independent.
							\item[(ii)] $\Ob_i = \Span\left\{\mathbf{e}_j|O_i{'} \mathbf{e}_j\neq \mathbf{0},j\in\Jc\right\}.$
						\end{enumerate}
					Due to space limit, we omit the proof that (i) and (ii) are equivalent. Since (i) is proved in \cite{zishuo_RNC2021} Appendix A Lemma 4, we conclude that (ii) holds true and thus the proof is completed.
					$\qed$
				\end{pf}
			
			\section{Proof of Lemma \ref{lm:subspace}}   \label{ap:proof_lm_subspace}
			We first prove equation \eqref{eq:HAH}. Define the $j$-th row of ${A}$ as $A_j$, then
			$$H_i A=\begin{bmatrix}
					A_{j_1}\\
					A_{j_2}\\
					\vdots\\
					A_{j_{n_i}}\
				\end{bmatrix},$$
			where $j_1,j_2,\cdots,j_{n_i}$ is the elements in $\Qc_i$.
			Noticing that $A$ is in the Jordan canonical form, for each $j\notin \Qc_i$, the $j$-th column of $H_iA$ is a zero column. As a result, 
			\begin{align*}
					\tAi H_i=&H_i A H'_i H_i=H_i A \diag\left(\sum_{k\in\Qc_i} \mathbf{e}_k \right)=H_i A .
				\end{align*}
			Similarly, according to the Jordan canonical form, the $j$-th element of row vector $C_i$ is zero if $j\in\Qc_i$. Thus, equation \eqref{eq:CHH} holds true due to
			\begin{align*}
					\tCi H_i=& C_i H'_i H_i=C_i \diag\left(\sum_{k\in\Qc_i} \mathbf{e}_k \right)=C_i.
				\end{align*}
			We proceed to prove that system $(\tAi,\tCi)$ is observable. The corresponding observability matrix is 
			\begin{align*}
					\tilde{O}_i &=\begin{bmatrix}
							\tCi\\
							\tCi \tAi\\
							\vdots \\
							\tCi \tAi^{n-1}
						\end{bmatrix}=
					\begin{bmatrix}
							C_iH_i'\\
							C_iA H_i'\\
							\vdots \\
							C_iA^{n-1}H_i'
						\end{bmatrix}\\
					&=O_iH_i'=\begin{bmatrix}
							O_i\mathbf{e}_{j_1} & O_i\mathbf{e}_{j_2}&\cdots &O_i\mathbf{e}_{j_{n_i}}
						\end{bmatrix}
				\end{align*}
			where $\{j_2,j_2,\cdots,j_{n_i}\}=\Qc_i$. Since it is proved in \cite{zishuo_RNC2021} Appendix A Lemma 4 that vectors in set $\{O_i \mathbf{e}_j |j\in\Qc_i\}$ are linear independent, we know that $\tilde{O}_i$ is full column rank. Thus $\rank(\tilde{O}_i)=n_i$ and is equal to state dimension of system $(\tAi,\tCi)$, which means that the sub-system is observable. $\qed$ 
		\end{document}